\documentclass[mathleft
]{an}
\usepackage{graphicx}
\usepackage{times}
\begin{document}

% The following seven commands are intended for editorial usage and should be ignored by
% the author(s).
\Pagespan{1}{}% Document's page range. 
% If second parameter is left empty, the last page is computed automatically.
\Yearpublication{2012}%
\Yearsubmission{2012}%
\Month{1}%   
\Volume{1}%  
\Issue{1}% 
% \DOI{This.is/not.aDOI}% 

\title{X-ray emission from RX~J1720.1+2638 and Abell~267: a comparison between a fossil and a  non-fossil system}
\author{Elena Jim\'enez-Bail\'on\inst{1}\fnmsep\thanks{Corresponding author:
  \email{elena@astro.unam.mx}\newline}
%Example 
%for footnote, note the usage of the \texttt{fnmsep}
%command as separator between institute number and footnote mark} 
\and  M\'onica Lozada-Mu\~noz\inst{1}
\and  J. Alfonso L. Aguerri\inst{2,} \inst{3}
}

\titlerunning{X-ray emission from RX~J1720.1+2638 and Abell~267}
\authorrunning{Jim\'enez-Bail\'on, Lozada-Mu\~noz \& J.A.L. Aguerri}
\institute{
Instituto de Astronom\'{\i}a
Apdo. 70-264,
 Cd. Universitaria, 
M\'exico DF 04510, 
M\'exico
\and 
Instituto de Astrof\'{\i}sica de Canarias
c/ V\'{\i}a L\'actea, s/n 
E38205 La Laguna, Tenerife, Spain
\and
Departamento de Astrofisica 
Universidad de La Laguna 
E-38205 La Laguna, Tenerife, Spain}

\received{}
\accepted{}
\publonline{later}
\keywords{galaxies: clusters: individual: RX~J1720.1+2638, Abell~267, X-rays: galaxies: clusters}

\abstract{%
  We present the XMM-Newton X-ray analysis of RX~J1720.1+2638 and Abell~267, a non-fossil and a fossil system, respectively.  The whole spectrum of both objects can be explained by  thermal emission. The luminosities found for   RX~J1720.1+2638 and Abell~267 in the 2-10~keV band are  6.20$^{+0.04}_{-0.02}\times10^{44}$ and 3.90$^{+0.10}_{-0.11}\times10^{44}\,$erg$\,$s$^{-1}$, respectively. The radial profiles show  a cool core nature for  the non-fossil system RX~J1720.1+2638,  while Abell~267 shows a constant behaviour of temperature with radius.  Metallicity profiles have also been produced, but no evidence of any gradient was detected due to the large uncertainties in the determination of this parameter. Finally, density and mass profiles were also produced allowing to derive M$_{500}$ for RX~J1720.1+2638 and Abell~267. The masses obtained are high, in the range of $(5-7)10^{14}\,M_{\odot}$. The X-ray properties obtained for both systems are not always in good agreement with what is expected: cool cores are expected for fossil systems, as Abell~267, considering them as and relaxed systems. However, the decrement of the temperature in Abell~267 could start at lower radii. Also the presence of a recent merger in Abell~267, already suggested in the literature,  could have  increased the central temperature.  The non-fossil system RX~J1720.1+2638 actually exhibits a cool core profile, but also evidence of a recent merger has been reported.}

\maketitle

\section{Introduction}

\begin{table*}[hat]
\footnotesize
 \centering%%%
\caption{Values of the parameters for best fit deprojected models and goodness of fit for all annuli spectra extracted in RX~J1720.1+2638 and Abell~267.}
\label{data}

\begin{tabular}{cccccccccc}\hline
\\
\multicolumn{4}{c} {\bf RX~J1720.1+2638}  & \multicolumn{4}{c} {\bf Abell 267}  \\
Radius &  kT & Z & $\chi^2_\nu$  & D.O.F.& Radius &  kT & Z & $\chi^2_\nu$  & D.O.F.\\ 
(kpc) & (keV) & (Z$_\odot$) &  & & (kpc) & (keV) & (Z$_\odot$) \\
\hline
0-27         &  $4.08\pm0.18$	         &  $0.52^{+0.10}_{-0.09}$		& 1.24 & 230 	& 0-73          &  $6.8^{+1.3}_{-0.4}$	         &  $0.4\pm0.3$ 			& 1.20 & 130 \\
27-54       &  $4.46\pm0.15$	         &  $0.58^{+0.09}_{-0.08}$		& 1.30& 274 	& 73-110      &  $6.2^{+1.5}_{-1.1}$	         &   $\le0.4$			& 0.91  	& 126 \\
54-82       &  $4.7^{+0.4}_{-0.2}$  	&  $0.50\pm0.11$			& 1.32 & 270  	& 110-147   &  $5.8^{+1.2}_{-0.4}$ 	         &   $0.4^{+0.4}_{-0.3}$ 	&1.00 & 124 \\
82-110     &  $5.4\pm0.4$	                   &  $0.41\pm0.14$			& 1.22 & 264  	& 147-220   &  $5.4\pm0.6$			&  $0.30^{+0.19}_{-0.17}$ & 1.23 & 163 \\
110-137   &  $5.8^{+0.5}_{-0.4}$	&  $0.31^{+0.13}_{-0.13}$		& 1.18 & 252  	& 220-331   &  $6.7\pm0.7$			&  $0.33\pm0.16$		& 1.16 & 183 \\
137-191   &  $6.4\pm0.4$			&  $0.31\pm0.09$	                   & 1.07 & 308 \\
191-273   &  $6.7\pm0.5$			&  $0.25\pm0.10$	                   & 1.13& 332 \\
273-410   &  $7.4^{+0.8}_{-0.6}$	&  $0.34^{+0.14}_{-0.13}$		 & 1.03 & 368 \\

\hline
\end{tabular}
\end{table*}

Fossil systems (Harrison et al. 2012) are systems of galaxies dominated by a single and isolated massive elliptical galaxy. The classification as a fossil system requires a gap in the R-band magnitude of  two or more for the two brightest galaxy in half of the viral radius and extended X-ray emission, L$\ge1\times10^{42}\,h_{50}^{-2}$~erg$\,$s$^{-1}$ (Jones et. al 2003). The current theory suggests that these objects collapsed in the early Universe with enough time to be able to merge the more massive galaxies (Jones et al. 2003;  Khosroshahi et al. 2004; Khosroshahi et al. 2007) and that are the most undisrupted systems of galaxies.  This scenario is supported by both observational studies (Harrison et al. 2012; Jones et al. 2003; Khosroshahi et al. 2004; Khosroshahi et al. 2007) and cosmological simulations (D$'$Onghia et al. 2005; von Benda-Beckmann et al. 2008). Alternatively, other studies show that these systems actually present a deficit of L$_*$ galaxies (Aguerri et al. 2011; Mendes de Oliveira et al. 2006; Mulchaey \& Zabludoff 1999).  According to both observations and simulations, there is therefore no clear scenario for the physical origin of fossils. Here, we present the EPIC XMM-Newton data analysis of the two clusters RX~J1720.1+2638 and Abell~267 selected from Santos et al. 2007 and classified  as fossil systems of galaxies. A detailed analysis of optical data (J. L. Aguerri, private communication) showed that  RX~J1720.1+2638 cannot be classified as a fossil system according to its optical photometric and spectroscopic properties. However the general properties of both systems (i.e. $\sigma\sim1100$~km$\,$s$^{-1}$ for both systems, Girardi et al 2013, in preparation) are very similar and therefore this study consists of a good comparison between a fossil and a non-fossil system.

\section{Data reduction and analysis}

The XMM-Newton  observations of  RX~J1720.1+2638 (ObsID:0500670201) and Abell~267 (ObsID: 0084230401) were processed using  SAS, v11.0  and using the most updated calibration files available in November 2011. Event lists from EPIC detectors were filtered to ignore periods of high background flaring following Piconcelli et al. (2004).The net exposures for RX~J1720.1+2638 and Abell~267 are 23.4 and 12.7 ks, respectively. Background spectra were extracted from blank sky event files, provided by the XMM-Newton EPIC Background Blank Sky team (Carter \& Read 2007). The  blank sky event files were requested using the same criteria of the observations (filter, mode). Source and background spectra, along with associated response matrices and ancillary response files were obtained with SAS. We simultaneously fitted the pn and MOS spectra using
Xspec v12.7.1.  The spectral analysis was performed in the 0.3-8 keV band  with a Hubble constant of 70 kms$^{-1}$Mpc$^{-1}$ and $\Omega_M=0.3$ and $\Omega_{\Delta}=0.7$. The redshift of the objects are 0.159 and 0.231 for RX~J1720.1+2638 and Abell~267, respectively.  These values correspond to a distance of D$_A$=565.9 Mpc  and  1 arcsec=2.732 kpc and D$_A$=757.9 Mpc and 1 arcsec=3.674 kpc for RX~J1720.1+2638 and Abell~267, respectively. Galactic absorption has been taken into account for both objects: N$_H =3.6\times10^{20}cm^{-2}$ for  RX~J1720.1+2638 and N$_H =2.75\times10^{20}cm^{-2}$ Abell~267 (Kalberla et al. 2005). In order to be able to perform $\chi^2$ technique to find the best fit model to our spectra, we grouped them in order to have at least 25 counts per channel.

 \begin{figure}
\centering
\includegraphics[width=50mm,angle=0]{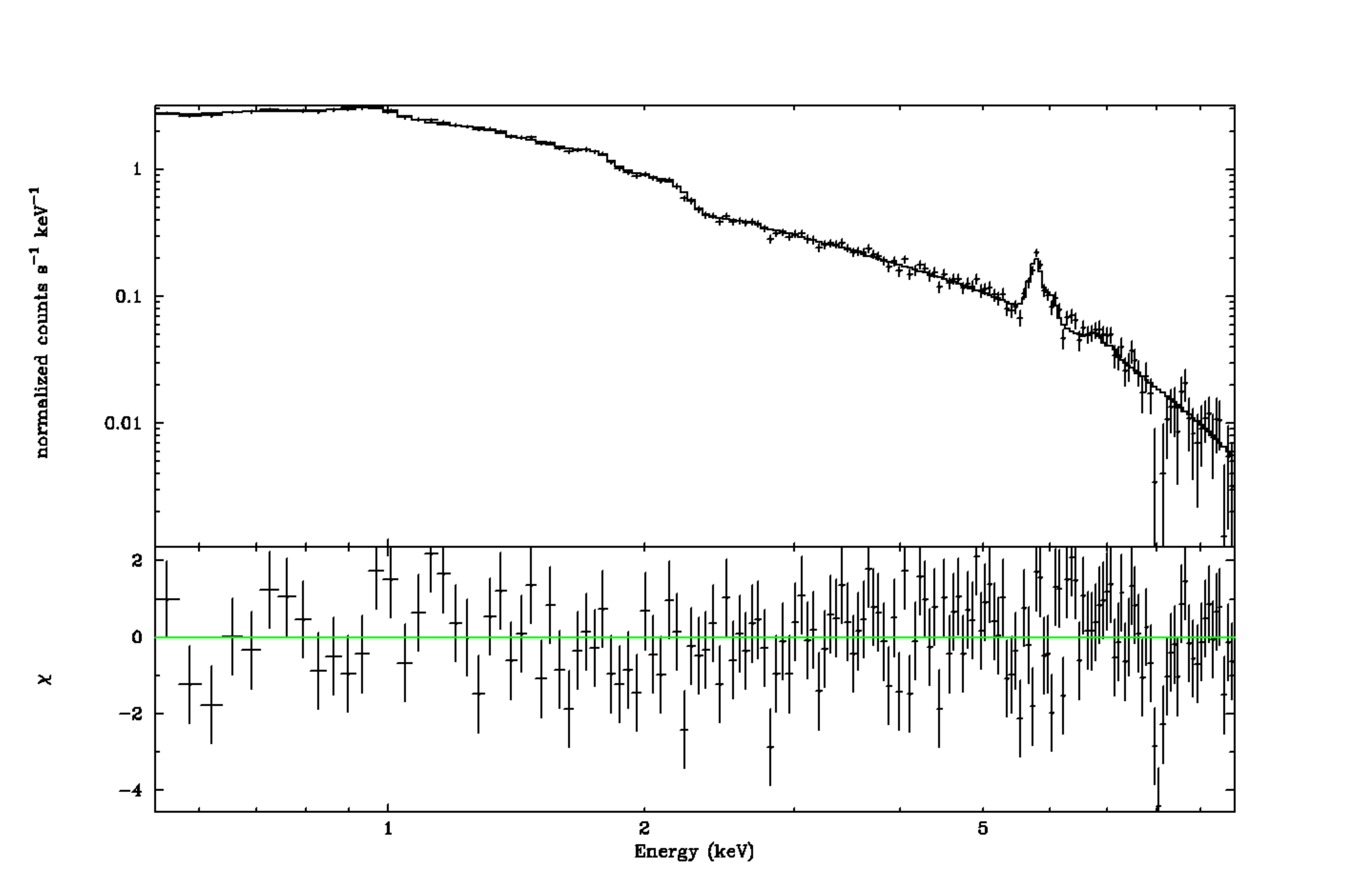}
\includegraphics[width=50mm,angle=0]{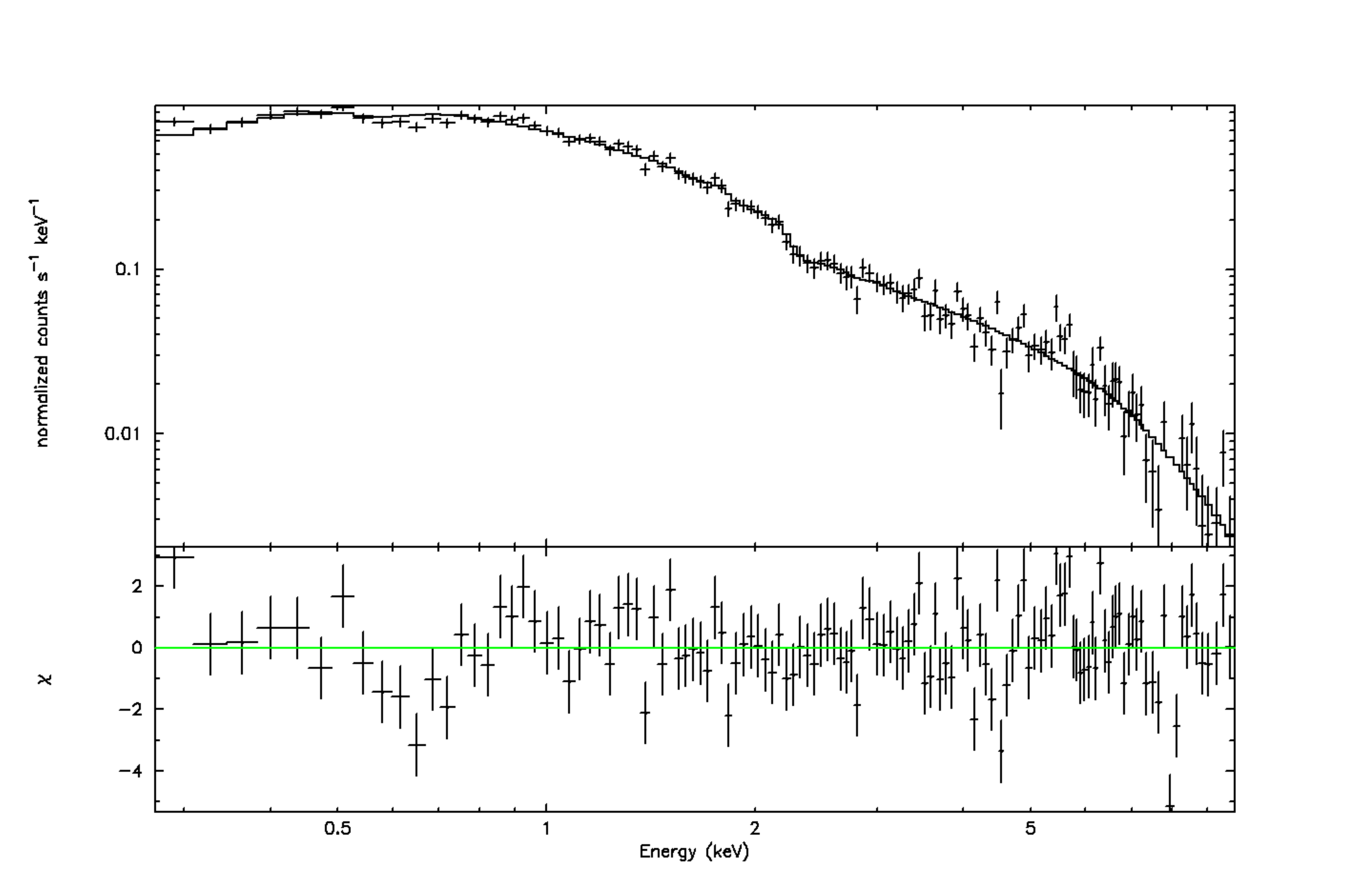}
\caption{PN spectra, best fit model and residuals for the whole emission of RX~J1720.1+2638 (up) and Abell~267(down). }
%For clarity, only EPIC-pn data are shown.}
\label{spectrum_fg31}
\end{figure}

%\begin{figure}
%\centering
%\includegraphics[width=62mm,angle=0]{spectrum_abell267c.pdf}
%\caption{Observed spectra, best fit model and residuals for the whole emission of   Abell~267.  For clarity, only EPIC-pn data are shown.}
%\label{spectrum_abell267}
%\end{figure}

The bulk of the emission of RX~J1720.1+2638 and Abell~267 extends to radii of  100 arcsec (275 kpc) and 90 arcsec (330 kpc), respectively. The spectra extracted to investigate the whole emission of the clusters  can be explained by a thermal emission in both cases.  The RX~J1720.1+2638 global emission was properly fitted with a double {\it mekal} model with temperatures of  kT$_{high} =6.8^{+0.5}_{-1.3}$~keV and kT$_{low} =2.3^{+0.8}_{-0.3}$~keV  and metallicities of $0.46^{+0.07}_{-0.05}$ and $0.25^{+0.17}_{-0.08}$ solar times, respectively. The goodness of the fit was $\chi^2$/dof=484/397. The absorbed fluxes measured in the 0.5-2 (2-10) keV bands were $5.22\pm0.02(8.56^{+0.06}_{-0.03})\times10^{-12}$ erg$\,$cm$^{-2}\,$s$^{-1}$ and the corresponding unabsorbed  luminosities were 3.97$\pm0.02(6.20^{+0.04}_{-0.02})\times10^{44}\,$erg$\,$s$^{-1}$. The higher temperature component accounts for 70\% of the bulk of the soft emission and up to 90\% goes to hard emission. Abell~267 spectrum  was satisfactorily fitted with two thermal components, kT$_{high} =11^{+10}_{-8}$~keV and kT$_{low} =1.4\pm0.3$~keV and associated metallicities of Z$_{high} \le0.3$ and Z$_{low} =0.05\pm0.4$ times the solar metallicity.  The  goodness of the fit was $\chi^2$/dof=416/292. The absorbed fluxes measured in the 0.5-2(2-10) keV bands were $1.40\pm0.02(2.44^{+0.06}_{-0.10})\times10^{-12}$ erg$\,$cm$^{-2}\,$s$^{-1}$ and the corresponding unabsorbed luminosities were $2.27\pm0.03(3.90^{+0.10}_{-0.11})\times10^{44}\,$erg$\,$s$^{-1}$. The higher temperature component accounts for 70\% of the bulk of the soft emission and up to 95\% of the hard emission. No absorption above the Galactic value has been found for any of the two spectra.  Figure~\ref{spectrum_fg31} shows the observed spectra, the best fit model, and the residuals for RX~J1720.1+2638 and Abell~267.

 \begin{figure*}
 \centering
\includegraphics[width=48mm]{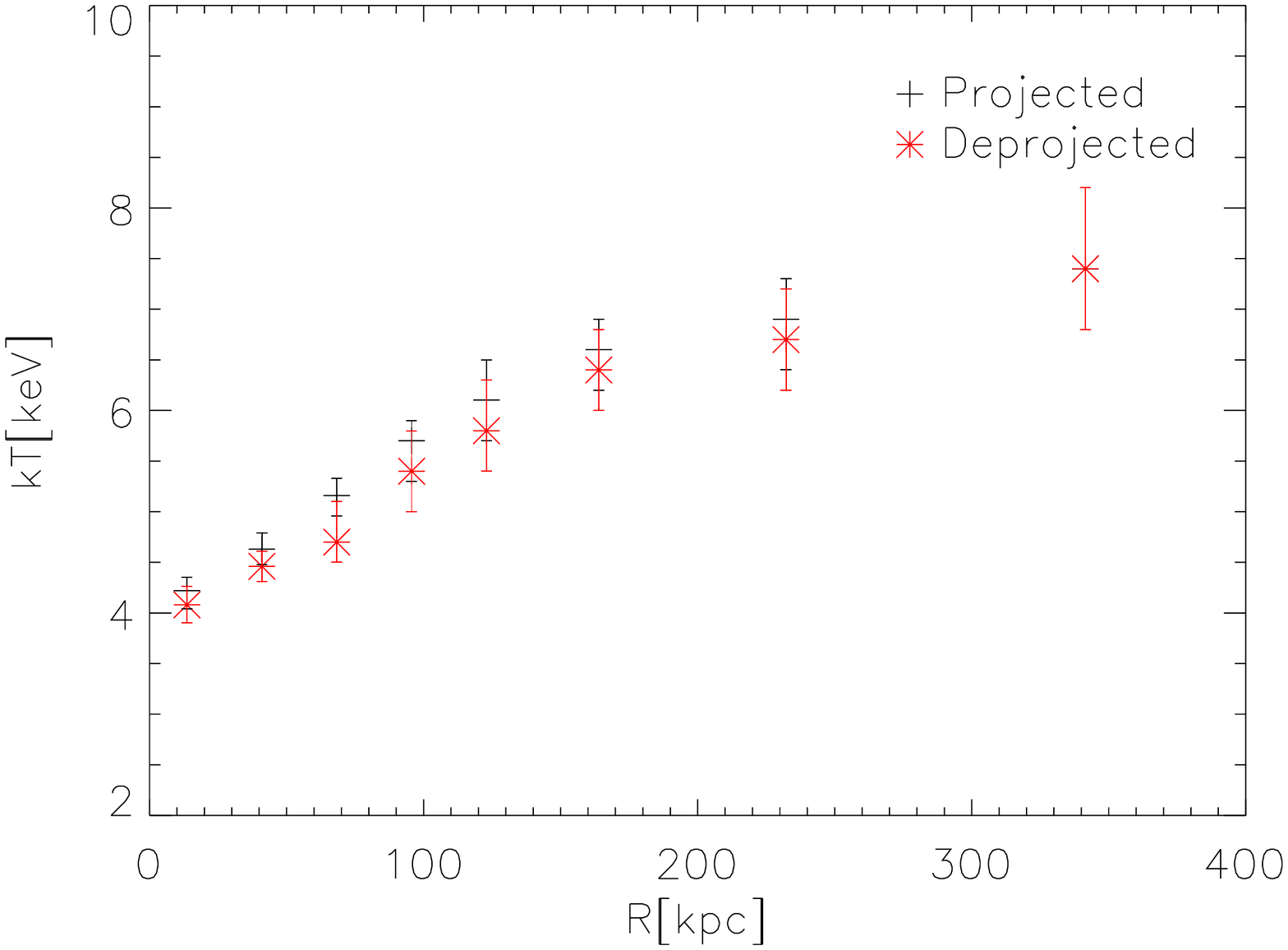}\includegraphics[width=50mm]{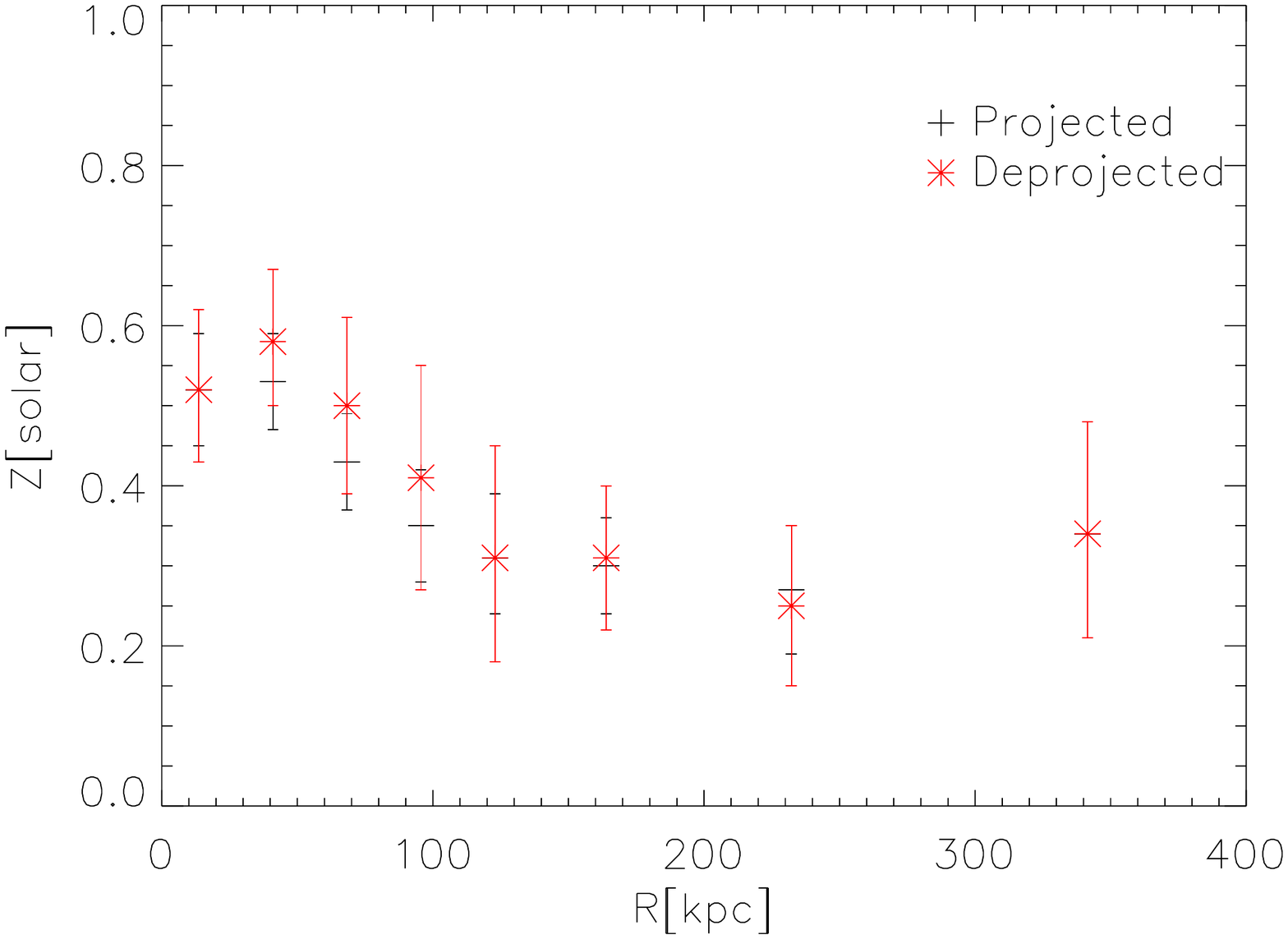}\\
\includegraphics[width=48mm]{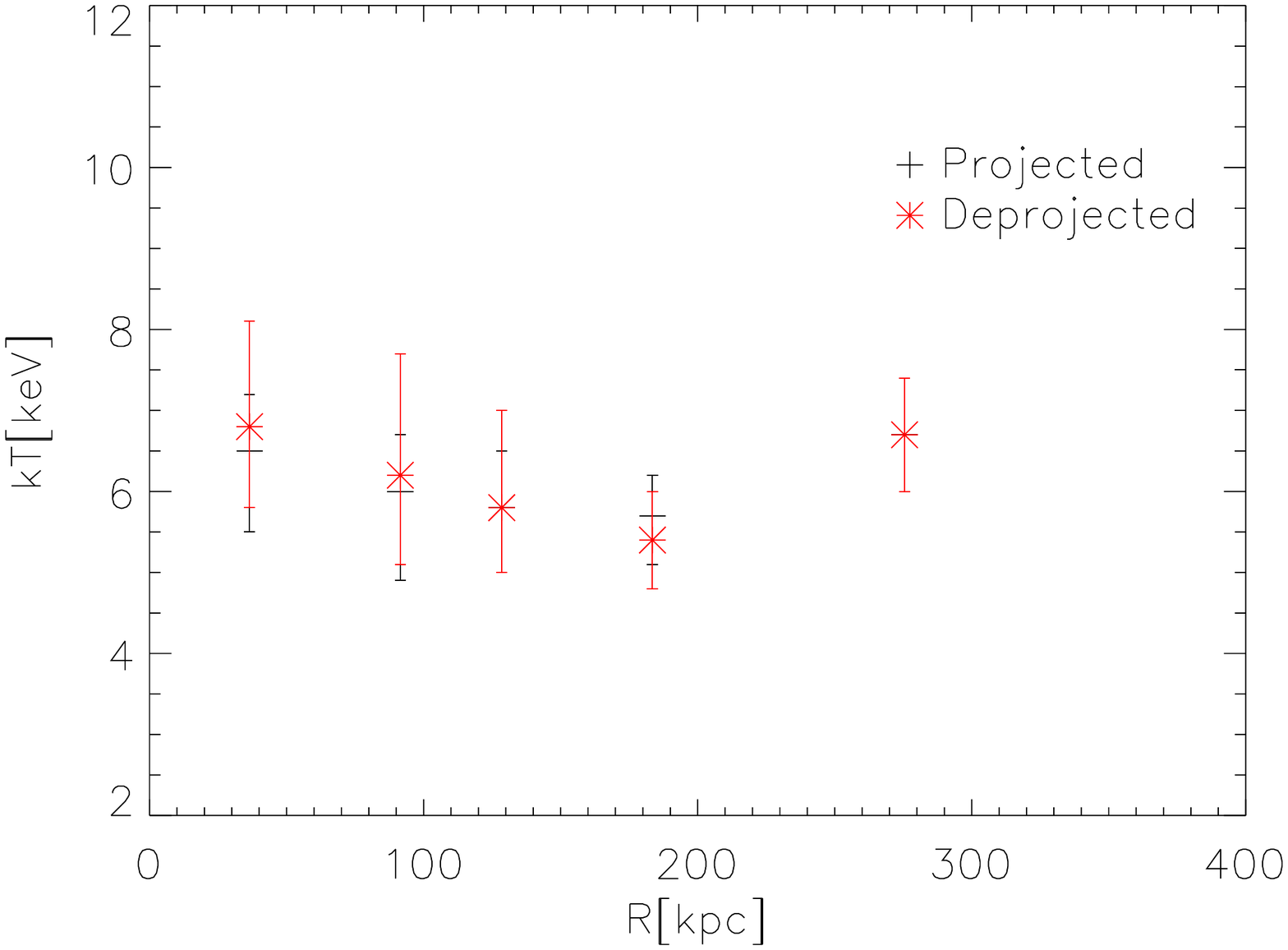}\includegraphics[width=48mm]{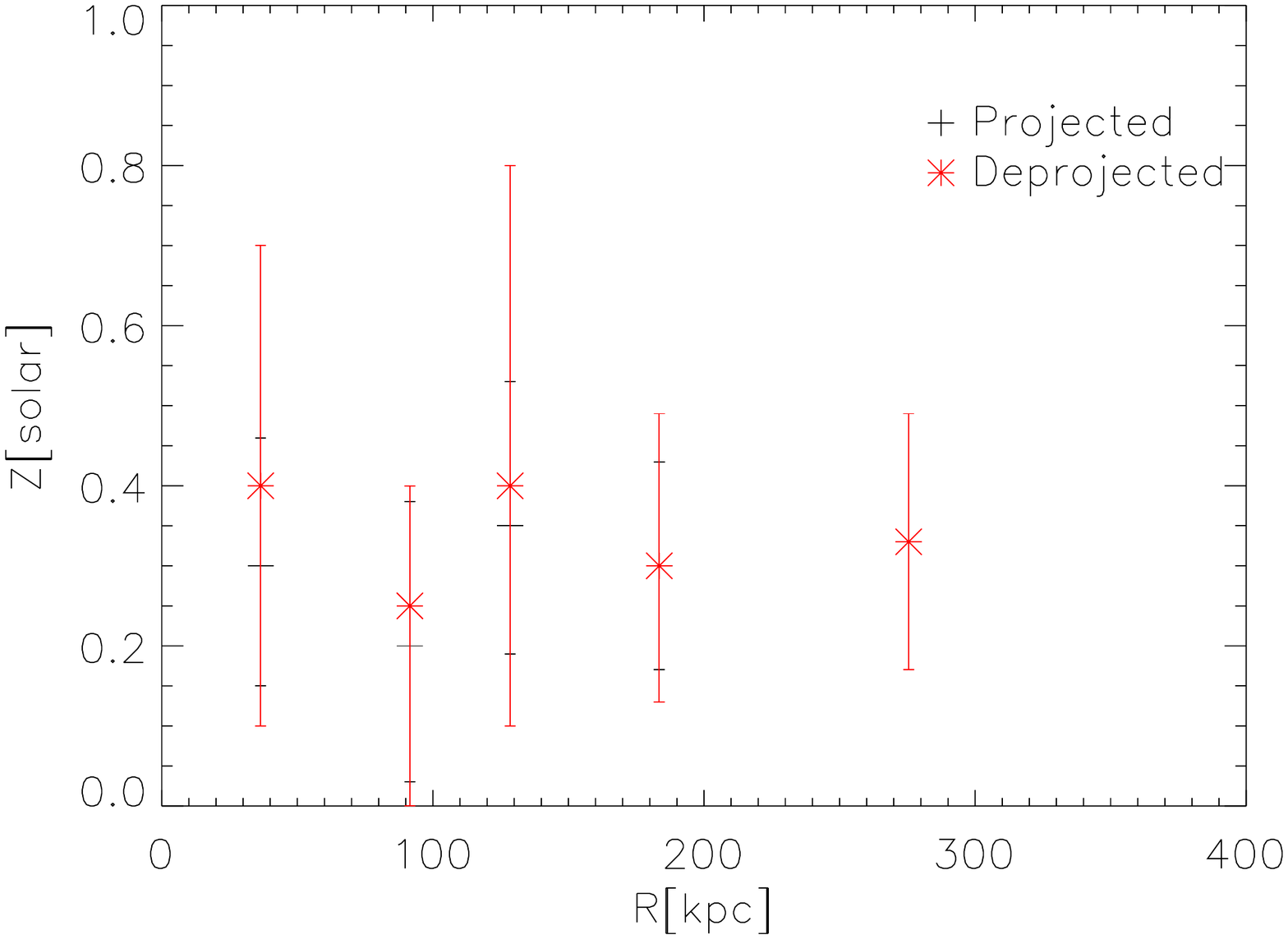}

\caption{Projected  ($+$)   and deprojected ($\ast$) temperature profile  and  metallicity of RX~J1720.1+2638 (up) and  Abell~267(down). }
\label{profile_temp_fg31}
\end{figure*}

 The signal-to-noise of the data allows  a radial spectral analysis for both sources. We therefore extracted the spectra for annular regions from the highest emission peak to the maximum extension radius. In order to be able to perform a spectral analysis with enough signal-to-noise to derive well-defined parameters, we extracted annular regions with at least 2000 counts, after background subtraction for Abell~267: five annuli have been extracted. For RX~J1720.1+2638, we extracted the same annuli  as in Mazzotta et al. (2001) in which the authors analysed Chandra data (obsid 1453) to allow comparison. Table~\ref{data} shows for both sources  the inner and outer radius of the annuli extracted and the parameters of the fitted model, {\it zwabs$\cdot$mekal} for Abel~267 and  {\it zwabs$\cdot$raymond-smith} (for comparison with Mazzota et al. 2001 work). In order to obtain the properties of the sources in a 3D space from the 2D spectrum projected, we use the deprojection technique  known as {\it onion peeling} (Ettori 2002).  Goodness of the fits are also shown in Table~\ref{data}. Using these values, we obtained temperature and metallicity profiles for each source in order to investigate the presence of a cool core and/or a metallicity gradient. In this sense, we found that while in Abell~267 the temperature remains constant at all radii, the temperature decreases to the centre of cluster RX~J1720.1+2638, as it can be seen in Figure~\ref{profile_temp_fg31} (left panel). Unfortunately, the accuracy on the determination of the metallicities is poor and therefore nothing can be said about the radial behaviour of this quantity (see Figure~\ref{profile_temp_fg31}, right panel).  However, a subtle increment towards the inner radii can be appreciated in the metallicity profile of RX~J1720.1+2638. The deprojected temperatures found for RX~J1720.1+2638 are compatible with those found by Mazzotta et al. (2001) from the Chandra observation (obsid 1453) only for annuli 1 to 5. For the last three annuli, the temperatures found by Mazzotta et al. (2001) are higher by 2-3keV. We re-analysed the Chandra data  following the standard method for extended sources of this observation and another available in the archive, obsid. 4361. We find deprojected temperatures lower but compatible with those of Mazzotta et al. (2001) for all annuli. In particular, XMM-Newton deprojected temperatures are fully compatible with those obtained from Chandra observation 4361, except for the last point. We inspected the XMM-Newton spectrum of this last annulus and we noticed a very large background emission above 2 keV which we suspect is responsible of mimicking the measured low temperature. We therefore decided to remove in the following the last point from our further analysis.
   
\begin{figure}
\centering
\includegraphics[width=58mm]{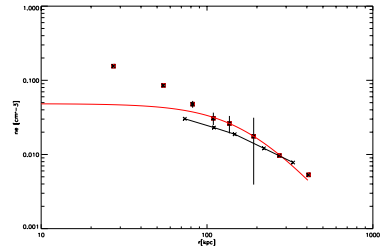}
\caption{Electron density profile and fitted $\beta$-model for Abell 267 (black) and RX~J1720.1+2638 (red). }
%The fit in RX~J1720.1+2638 has only been applied for the external part of the systems due to the artificial increment of the electron density in the central parts typically observed in cool core systems.}
\label{density_fg31}
\end{figure}

  Based on the definition of the normalisation of the thermal emission model, it is possible to determine the electron density of the media. Therefore, a density profile has also been constructed for both sources. Figure~\ref{density_fg31}  shows the obtained density profiles for RX~J1720.1+2638 and Abell~267. Assuming spherical symmetry, we fitted  a $\beta$-model to the density profile.  The fit allowed us to determine the mean central density and the radius of the nucleus: $n_0$=0.05~cm$^{-3}$, $r_0$=175~kpc  and $\beta=0.84$ for RX~J1720.1+2638 and  $n_0$=0.043~cm$^{-3}$, $r_0$=80~kpc and $\beta=0.38$ for Abell~267. It is also possible to derive the mass profile of the systems by assuming hydrostatic equilibrium, a spheric mass distribution, constant temperature of the intracluster gas, and a NFW (Navarro et al. 1995) profile for the distribution of the density $\rho_m$ of the media. In particular, we derived for  RX~J1720.1+2638 and Abell~267 the values of M$_{500}$ and r$_{500}$. We obtained M$_{500}=(7\pm2)\times10^{14}\,M_{\odot}$ and r$_{500}=900\pm4100$~kpc for the first and for Abell~267, we found that M$_{500}=(4.7\pm1.8)\times10^{14}\,M_{\odot}$ and r$_{500}=780\pm400$~kpc.  The mass of Abell~267 found is compatible with the value reported in Zhang et al. (2008) within 2$-\sigma$. We have also used the M-T relationships in Chen et al. 2007 and Finoguenov et al. 2001 to derive the masses of the systems. For Abell~267 the calculated M$_{500} $ range (7-9)$\times10^{14}\,M_{\odot}$ is fully compatible with what was found using the NFW profile. For RX~J1720.1+2638, excluding the cool core annuli, assuming to be from 1-4, and the last annulus, we obtain masses varying in the range of (6-8)$\times10^{14}\,M_{\odot}$, in good agreement with the results form the NFW profile.

% In order to better understand the thermal history of the cluster  (change on the amount of kinetic energy of the intracluster gas) it is useful to study the behavior of the entropy, $S=kT/n_e^{2/3}$.

\section{Results and Conclusions}

Here in this work, we analysed the  EPIC XMM-Newton data of the fossil system Abell~267 and the non-fossil system RX~J1720.1+2638. Firstly, global X-ray properties have been derived for both systems. The integrated spectra of both sources can be explained by pure thermal emission with mean temperatures of $\sim$5-6~keV for both sources. This values are typical of massive clusters of galaxies. Metallicities range from 0.1 to 0.5 Z$_\odot$, also compatible with what was found by Balestra et al. 2007 for a sample of 56 clusters at these distances observed with Chandra. The luminosities found for   RX~J1720.1+2638 and Abell~267 in the 2-10~keV band are 6.20$^{+0.04}_{-0.02}\times10^{44}$ and 3.90$^{+0.10}_{-0.11}\times10^{44}\,$erg$\,$s$^{-1}$, respectively. Fossil systems show an excess in X-ray luminosity of about one order of magnitude compared to non-fossil systems for a given total optical luminosity (Jones et al. 2003; Khosroshahi et al. 2007; see also Harrison et al. 2012). The observed R-band magnitudes for RX~J1720.1+2638 and Abell~267 are very similar, -24.3 and -24.9, respectively. According to this result, we would expect a higher X-ray  luminosity for Abell~267 than for RX~J1720.1+2638 but interestingly the measured values are of the same order of magnitude. Moreover, the luminosities of the two clusters scale with their mass as expected. A relatively boosted luminosity of Abell~267, due to its cooling flow (e.g. Chen et al. 2007) is not found by our analysis.

Radial profiles of the temperature and metallicity were also calculated for both sources. The non-fossil system, RX~J1720.1+2638, shows a decrement in temperature for inner radius, visible below 0.1r$_{200}$ ($\sim0.18r_{500}$ calculated to be at around 160~kpc). This value is in good agreement with what was found for a sample of 15 nearby clusters observed with XMM-Newton (Pratt et al. 2007). A similar decrement was found by Mazzotta et al. (2001) using Chandra data. For Abell~267, the temperature  remains constant within the errors for all radii.  In this sense, and assuming the theory in which fossil systems are considered as the end product of galaxy merging, then no recent merger could have occurred (on average, only one galaxy has been accreted since z$\sim1$, von Benda- Beckman et al. 2007). This, in term, translates to an absence of any heating source to prevent the decrement in temperature at inner regions of the fossil system. Due to this relaxed nature of fossil systems, the presence of cool cores is expected. However, the non-detection of a cool core in Abell~267 could be due to several causes. One possibility is that for some reason, the decrement in temperature for fossil systems begins at lower radii than in normal clusters, i.e. $\le0.1r_{200}$. Unfortunately, this possibility could not be tested due to the limited signal-to-noise of our data. Another possibility is that the core of the system is being heated. One alternative is that the source of heating is  an AGN. However, no evidence of any hidden AGN was detected. One other possibility is that  the heating could be due to a recent major merger. Zhang et al. 2008 stated that Abel~267 has actually a disrupted morphology. Mazzotta et al. (2001) also suggest the presence of a merger in RX~J1720.1+2638, which supports  the non-fossil nature of this system. We also produced the electron density and mass profiles for both systems. The derived values of M$_{500}$ are of the order of $(5-8)\times10^{14}\,M_{\odot}$. Proctor et al. (2011) found that fossil systems have masses comparable to those of clusters, and Harrison et al. 2012 showed that the brightest galaxy of fossils are among the most massive galaxies in the Universe. However, even taking into account these considerations the values found for the masses of the systems are very high. Moreover, both sources present evidence of a recent merger, and therefore our  assumption of spherical symmetry could be introducing uncertainties in the mass determinations.

In summary, we presented the results on the analysis of two clusters: RX~J1720.1+2638 classified as a non-fossil system with the presence a cool core but interestingly with evidence of a recent merger; and the fossil system Abell~267 with no evidence of cool core. The results obtained for both objects are therefore unexpected for their nature and an evidence  that the current scenario for fossil systems as  relaxed systems, end products of galaxy mergers is not directly applicable to all fossil systems, as it is the well-studied case of Abell~267. 
%Further X-ray analysis of large and well-exposed samples are mandatory to properly determined the common properties fossil systems and therefore to be able to seed light on the physical origin of these peculiar systems.

%\begin{figure}
%\centering
%\includegraphics[width=58mm]{FG2_ne057kev.pdf}
%\caption{Electron  density profile and fitted $\beta$-model for Abell~267.}
%\label{density_fg2}
%\end{figure}

%\begin{figure}
%\includegraphics[width=50mm,height=28mm]{testbild.eps}
%\caption{The caption of a figure shall describe the content of the figure.
%An exception is granted only to the Editor and its ghostwriter, who
%demonstrate the usage of this command.}
%\label{label1}
%\end{figure}

%A large figure spanning significantly more than one column shall be embraced by
%the  \verb+\begin{figure*}+ and \verb+\end{figure*}+ commands.

%\begin{table}
% \centering%%%
%\caption{Example of a well organized table}
%\label{tlab}

%\begin{tabular}{cc}\hline
%Quantity 1 & Quantity 2\\ 
%(unit1) & (unit2) \\
%\hline
%1 & 2 \\
%3 & 4\\
%\hline
%\end{tabular}
%\end{table}

%%%
%%% -MWL- 2006-01-13 auf Verlagswunsch wieder altes Bibliographie-Format
%%% 
%\subsection{References}

\acknowledgements
E. Jim\'enez-Bail\'on and M\'onica Lozada acknowledge  financial support  from CONACYT grant 129204.

%\newpage%%%%%%%%%%%%%%%%%%%%%%%%%%%%%%%%%%%%%%%%%%%%%%%%%%%%%%

%\appendix

\end{document}